\documentclass[draft]{agujournal2019}
\usepackage{apacite}
\usepackage{url} 

\usepackage{ragged2e}
\usepackage{graphicx}
\usepackage{amssymb}
\usepackage{amsthm}
\usepackage{mathtools}
\usepackage{lineno}
\usepackage{mathrsfs}
\usepackage{booktabs, caption, makecell}
\usepackage{threeparttable}
\usepackage{algorithm}
\usepackage{algcompatible}
\usepackage{lipsum}
\usepackage[nameinlink,capitalise]{cleveref}

\theoremstyle{remark}
\newtheorem*{remark}{Remark}

\draftfalse

\journalname{Wiley}

\begin{document}

%
%


\title{Deep convolutional neural networks for uncertainty propagation in random fields}

%
%




\authors{Xihaier Luo \affil{1}, Ahsan Kareem \affil{1}}


\affiliation{1}{NatHaz Modeling Laboratory, University of Notre Dame, Notre Dame, IN, USA.}




\correspondingauthor{Xihaier Luo}{xluo1@nd.edu}




\justify

\begin{keypoints}
\item A surrogate model for quantifying the effect of spatially random attributes in coupled elliptic systems is developed.
\item A hierarchical modeling design using deep convolutional neural networks as the underlying components is proposed to improve the model efficiency in terms of training and deploying.
\item The efficacy of the proposed learning approach is demonstrated using a benchmark structural problem with a variety of high-dimensional mapping relationships: $\mathbb{R}^{4096} \rightarrow \mathbb{R}^{4096}, \mathbb{R}^{4096} \rightarrow \mathbb{R}^{12288}, \text{and } \mathbb{R}^{8192} \rightarrow \mathbb{R}^{8192}$.
\end{keypoints}

%
%


\begin{abstract}
    The development of a reliable and robust surrogate model is often constrained by the dimensionality of the problem. For a system with high-dimensional inputs/outputs (I/O), conventional approaches usually use a low-dimensional manifold to describe the high-dimensional system, where the I/O data is first reduced to more manageable dimensions and then the condensed representation is used for surrogate modeling. In this study, we present a new solution scheme for this type of problems based on a deep learning approach. The proposed surrogate is based on a particular network architecture, i.e. the convolutional neural networks. The surrogate architecture is designed in a hierarchical style containing three different levels of model structures, advancing the efficiency and effectiveness of the model in the aspect of training and deploying. To assess the model performance, we carry out uncertainty quantification in a continuum mechanics benchmark problem. Numerical results suggest the proposed model is capable of directly inferring a wide variety of I/O mapping relationships. Uncertainty analysis results obtained via the proposed surrogate have successfully characterized the statistical properties of the output fields compared to the Monte Carlo estimates.
\end{abstract}


%
%

%


%
%
%
%

\section{Introduction}
\label{sec1}
\begin{figure}[H]
\centering
\includegraphics[width=1.0\textwidth]{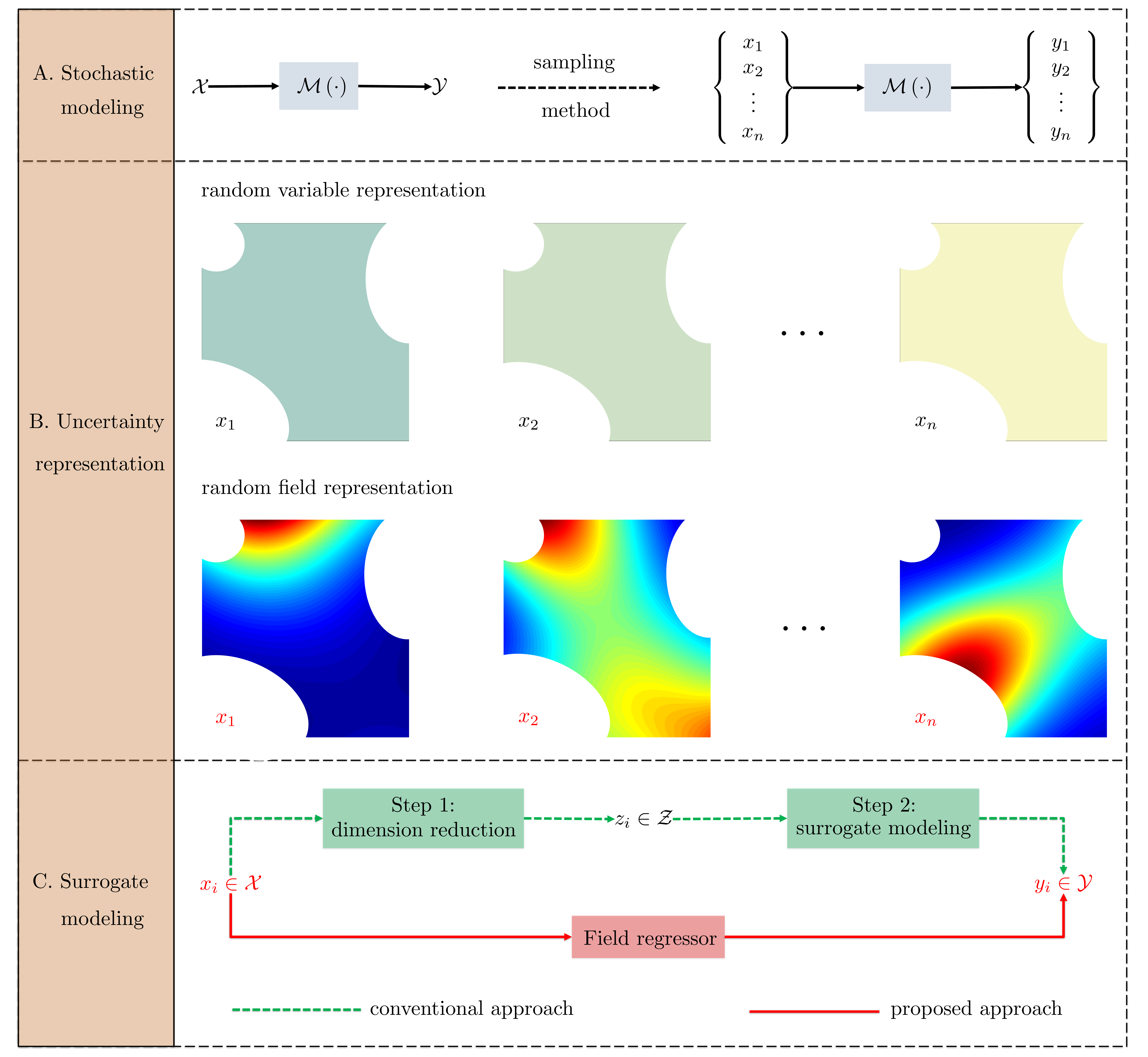}
\caption{A. Stochastic modeling using the Monte Carlo method; B. Random field theory to model the spatial variability; C. Surrogate modeling for random field uncertainty propagation} 
\label{fig: f1}
\end{figure}

The primitive parameters of real-world mechanical systems such as material properties and external excitations are often not known exactly and are subject to uncertainty. This uncertainty may arise from inevitable manufacturing imperfections, incomplete knowledge on system parametrization, and others. In most cases, uncertain parameters are modeled as random variables utilizing a sequence of probability distributions and the system of interest is characterized by stochastic partial differential equations (SPDEs) \cite{smith2013uncertainty}. Because the input parameters are random, the output of the system also becomes a random variable. To obtain comprehensive probabilistic descriptions for the random output quantities, one can rely on the ensemble of independent samples drawn from the distribution of the inputs using the Monte Carlo method, evaluating the system at a finite number of realizations and estimating the statistics of the target outputs (See \cref{fig: f1} .(A)). Such uncertainty quantification (UQ) task involving many system queries can be greatly expedited through the use of surrogate models especially for applications that depend on a time-demading simulator \cite{viana2014special}. 

Conceptually simple, a formidable challenge remains open concerning the development of a reliable and robust surrogate in cases where the input/output is no longer a random variable, instead, is a random field or even a set of random fields. For instance, one can treat material properties as random variables, which are random in essence but are uniformly distributed over the domain. Such treatment ignores the possible spatial variation and the realization of perfectly uniform material is unlikely to occur in practice, thus fails to provide a realistic reflection of the real-world situations \cite{babuvska2007stochastic, gurley1999applications, chen2016modeling}. On the other hand, the random field theory offers a more reliable avenue as it systematically takes the spatial correlation into accounts (See \cref{fig: f1} .(B)). The numerical discretization of a random field with small correlation can easily result in a large number of random variables. Unfortunately, most existing surrogate models dealing with uncertainty propagation tasks have difficulty scaling to high-dimensional problems and the conjunction use of the random field and the time-demanding computer model can be prohibitive \cite{schwab2011sparse, babuvska2007stochastic, dai2017wavelet}.

Given that the direct construction of surrogate models yielding an easy-to-evaluate mechanism between high-dimensional inputs and outputs can be extremely demanding, practitioners usually attempt to project the given random field data from its original high dimension to a lower dimension. The standard approach to performing dimensionality reduction on the continuous functional space in the UQ literature is the Karhunen-Lo\`eve expansion (KLE) \cite{xiu2002wiener, ghanem2003stochastic, le2010spectral}. Known under different names such as principal component analysis, proper orthogonal decomposition, etc, the KLE first projects a random field, which is infinite dimensional, to finite dimensions using eigenfunctions of the random field's covariance function as basis functions. Then, the KLE represents the random field as a linear combination of these orthogonal eigenfunctions and their corresponding uncorrelated coefficients, which are random variables. In spite of numerous applications in which the KLE based dimensionality reduction scheme circumvents the curse of dimensionality, the KLE by definition can only preserve the first and second order statistics, i.e., the mean and covariance of the random field. However, uncertainties arise from material heterogeneities or randomly distributed force fields usually exhibit nonlinear structures \cite{kareem2008numerical}. Moreover, the KLE is an unsupervised learning algorithm from a machine learning perspective \cite{murphy2012machine}, indicating it does not explicitly integrate the output information into the dimensionality reduction procedure and additional classification/regression model is needed to complete the surrogate modeling task (See the upper green route in \cref{fig: f1}. (C)).

In recent years, deep neural networks (DNNs) have gained increased popularity in representing the high-dimensional solutions of a wide class of partial differential equations \cite{chen2018neural, raissi2019physics, sirignano2018dgm, mo2019deep, zhu2018bayesian}. Compared to the conventional surrogate modeling approaches, DNNs tackle the curse of dimensionality by learning the latent representation through the use of a series of layers, which are connected by affine operations and nonlinear activations. Usually formed as a supervised learning problem, DNNs are capable of providing an end-to-end field-to-field surrogate model, thus avoiding the need to build a dimension reduction model (See the lower red route in \cref{fig: f1}. (C)). In this work, we attempt to deploy one of the most widely used network architectures, the convolutional neural networks (CNNs), to the UQ analysis of civil structures with spatially varying system properties. Within the realm of neural networks modeling, CNNs are the workhorse of image classification \cite{gao2018deep, krizhevsky2012imagenet}, object detection \cite{cha2017deep, gopalakrishnan2017deep}, computer vision \cite{yeum2015vision, yang2018automatic}, and signal processing \cite{rafiei2018novel}. This type of neural networks gains computational efficiency in dealing with data of a grid-like topology by extracting multi-scale features within the modeling process. This unique learning property triggers the use of CNNs for surrogate modeling of the complex nonlinear mapping between discretized random fields.

In this paper, a machine learning based structural analysis framework is proposed to statistically characterize the system responses under various spatially varying properties. We formulate our surrogate model in a hierichiacal form where CNNs are adopted as fundamental model components. Because CNNs are prone to overparameterization, additional network connections between nonadjacent convolutional layers are created in such a way that extracted features can be reused, thus reducing the model parameters and strengthening the machine learning process \cite{he2016deep, huang2017densely}. Moreover, the operations of deconvolution included in the CNNs can generate checkerboard artifacts, especially in the context of regression problems of high-resolution images. To address this issue, we make the joint use of the upsampling methods such as bicubic resizing and convolution operation in the decoding process instead of conventional deconvolution \cite{odena2016deconvolution, dong2015image}. To make the training process more efficient, we incorporate the annealing learning strategy in the setting of stochastic gradient descent and adopt batch normalization method to avoid data distortion \cite{robbins1951stochastic, ioffe2015batch}. The overall machine learning methodology is applied to a plate-like structure governed by a set of fourth-order elliptic partial differential equations. We demonstrate the flexibility and robustness of our proposed surrogate model by considering one-to-one ($\mathbb{R}^{4096} \rightarrow \mathbb{R}^{4096}$), one-to-many ($\mathbb{R}^{4096} \rightarrow \mathbb{R}^{12288}$), and many-to-many ($\mathbb{R}^{8192} \rightarrow \mathbb{R}^{8192}$) mappings, where each one represents a discretized high-dimensional random field.

In the rest of this paper, \cref{sec2} gives the problem statement for using deep neural networks to propagate uncertainties represented by random fields. \cref{sec3} provides a comprehensive guide in terms of building, training, and validating the proposed surrogate. In \cref{sec4}, a systematic case study on first identifying intrinsic relationships between discretized random fields and then applying the learned surrogate to the uncertainty analysis is presented. Lastly, \cref{sec5} summarizes the principal conclusions and discusses the future work.

\section{Problem statement}
\label{sec2}
This paper is focused on the development of surrogate models for effectively propagating and quantifying the influence of spatially varying uncertainties on structural behaviors in applied mechanics problems. To fulfill this objective, we consider general boundary value problems of the following elliptic PDEs form:

\begin{equation}
    \label{eq: 21}
    \mathcal{L}_{\boldsymbol{s}} \left(  \boldsymbol{\Theta}(\boldsymbol{s}), \boldsymbol{u}(\boldsymbol{s}) \right) = \boldsymbol{f}(\boldsymbol{s}) \quad \text{in} \quad \mathcal{S}
\end{equation}

where $\boldsymbol{s}$ is the cartesian coordinates, $\mathcal{L}_{\boldsymbol{s}}$ is a differential operator, $\boldsymbol{\Theta}$ represents an input property field encompassed in the constitutive equation, $\boldsymbol{u}$ denotes an unknown solution to the equation, $\boldsymbol{f}$ represents the source term that drives the system, and $\mathcal{S}$ is a Lipschitz smooth domain. To determine a unique solution, boundary conditions, i.e. the Dirichlet and Neumann boundary conditions are imposed on $\Gamma_{\mathrm{D}} \subset \partial \mathcal{S}$ and $\Gamma_{\mathrm{N}} \subset \partial \mathcal{S}$, respectively:

\begin{equation}
    \label{eq: 22}
    \begin{aligned}
        \boldsymbol{u}=g^{(D)} & \quad \text{at} \quad \partial \Gamma_{\mathrm{D}} \\ 
     \nabla \boldsymbol{u} \cdot \hat{n}=g^{(N)} & \quad \text{at} \quad \partial \Gamma_{\mathrm{N}}
     \end{aligned}
\end{equation}

with $\hat{n}$ denoting the unit normal on the boundary. The operator $\nabla$ is considered only with respect to (w.r.t.) the spatial variables $\boldsymbol{s}$. In a typical probabilistic mechanics problem, one often deals with systems subjected to stochastic excitations $\boldsymbol{f}(\boldsymbol{s})$, or themselves involve random parameters $\boldsymbol{\Theta}(\boldsymbol{s})$, or both. A standard uncertainty analysis framework for quantifying the influence of system uncertainties comprise three parts, that is, uncertainty representation, uncertainty propagation, and uncertainty quantification \cite{smith2013uncertainty}.

\begin{remark}[1]
Without loss of generality, we formulate the problem (\cref{eq: 21}) using a general elliptic operator $\mathcal{L}_{\boldsymbol{s}}$. In the context of uncertainty quantification of structural systems, the operator $\mathcal{L}_{\boldsymbol{s}}$ could be the momentum balance equation, which reads the mass density, body force, initial velocity, etc as the inputs and computes the corresponding displacement field. And our goal is to characterize and manage the spatial variability induced uncertainty effects on the system performance.
\end{remark}

\subsection{Step 1: random fields for uncertainty representation}
\label{sec21}
Let $(\Omega, \mathscr{S}, \mathbb{P})$ be a complete probability space, where $\Omega$ is the set of the elementary events $\omega$, $\mathscr{S}$ is an $\sigma$-algebra on the set $\Omega$, and $\mathbb{P}$ is a probability measure. We are interested in quantifying the uncertainty associated with the solution $\boldsymbol{u}$ that is due to the spatial variability of the inputs $\boldsymbol{\Theta}, \boldsymbol{f}$. To achieve the goal, we use random fields, which are second order stationary and statistically homogeneous, to represent physically relevant variability \cite{shinozuka1972digital}. In general, a random field $\boldsymbol{r} ( \boldsymbol{s}, \omega )$ is a manifold in the Hilbert space consisting of a family of random variables. Hence, the solution mapping of \cref{eq: 21} in a geometrically bounded regime with discretized dimension $d$ is expressed as:

\begin{equation}
    \label{eq: 23}
    \boldsymbol{u} : \mathcal{S} \times \Omega \longrightarrow \mathbb{R}^{d}
\end{equation}

The central work of the first \textit{uncertainty representation} step is to generate random field samples. Usually, the random field generation covers two parts, where the first part focuses on generating a set of uncorrelated random variables and the second part transfers these random variables into a correlated random field with prescribed statistical properties. There are several methods available for this purpose and matrix decomposition method is considered in this paper \cite{davis1987production, constantine2010random, kareem2008numerical}. Specifically, the realization of a random field with known properties is obtained by multiplying the lower triangular matrix computed from the Cholesky decomposition of the prescribed covariance matrix of $\boldsymbol{r} ( \boldsymbol{s}, \omega )$ by a random vector comprising a set of standard normal random variables.

\subsection{Step 2: deep neural networks for uncertainty propagation}
\label{sec22}
Given the fact that the input uncertainty is represented by means of random fields, the second step of the uncertainty analysis is to propagate the spatial variability to the outputs. To achieve this goal, deep neural networks are considered in this paper for two main reasons. First, the performance function $f \left( \cdot \right)$ of structural systems with spatially varying properties is highly complex and implicit in general. To approximate the function $f \left( \cdot \right)$, neural networks based surrogate is a strong candidate since it has been shown that a single hidden layer based neural networks are universal function approximators under mild conditions \cite{hornik1989multilayer}. Secondly, going deep allows the network structure smoothly reduces the high dimensionality $d$ arises from the discretization of the random field \cite{bengio2012practical}. 

Mathematically, the approach of using deep neural networks to propagate random fields represented uncertainties can be summarized as follows: (1) collect the training dataset $\mathcal{D} = \{ \boldsymbol{x}_i, \boldsymbol{y}_i \}_{i=1}^{n}$ by querying $f \left( \cdot \right)$ with $\boldsymbol{x}$ and $\boldsymbol{y}$ denoting the input and output respectively. For instance, $\boldsymbol{x}$ can be the system property field $\boldsymbol{\Theta}(\boldsymbol{s})$ or the external loading field $\boldsymbol{f}(\boldsymbol{s})$ or both $\boldsymbol{x} = [ \boldsymbol{\Theta}, \boldsymbol{f} ]$ , and $\boldsymbol{y}$ usually refers to the solution field $\boldsymbol{u}$. (2) train the surrogate model using $\mathcal{D} = \{ \boldsymbol{x}_i, \boldsymbol{y}_i \}_{i=1}^{n}$. The central goal is to establish a functional relationship $\hat{f} \left( \cdot \right)$ in such a way that discrepancies between $f \left( \cdot \right)$ and $\hat{f} \left( \cdot \right)$ are optimally minimized. Once $\hat{f} \left( \cdot \right)$ is available, one can immediately use it for the uncertainty quantification.

\subsection{Step 3: Monte Carlo method for uncertainty quantification}
\label{sec23}
With the aid of an efficient surrogate model, uncertainties concerning the inputs can be effectively propagated into the system outputs. Subsequently, the last \textit{uncertainty quantification} step centers on estimating the statistical properties of the output. The most widely used objective of a standard UQ problem is the characterization of the first two statistical moments, i.e., the mean,

\begin{equation}
    \label{eq: 24}
    \boldsymbol{\mu}_{f}=\int_{\Omega} f(\boldsymbol{x}) p(\boldsymbol{x}) \mathrm{d} \boldsymbol{x}
\end{equation}

and the variance,

\begin{equation}
    \label{eq: 25}
    \boldsymbol{\sigma}_{f}^{2}=\int_{\Omega} \left(f(\boldsymbol{x})-\boldsymbol{\mu}_{f}\right)^2 p(\boldsymbol{x}) \mathrm{d} \boldsymbol{x}
\end{equation}

In this paper, we are also interested in the computation of the probability density function (PDF):

\begin{equation}
    \label{eq: 26}
    p_{f}(\boldsymbol{y})=\int_{\Omega} \delta(\boldsymbol{y}-f(\boldsymbol{x})) p(\boldsymbol{x}) \mathrm{d} \boldsymbol{x}
\end{equation}

Since the function $f\left( \cdot \right)$ may not always be known in closed form for the problem under consideration, the sampling methods such as the Monte Carlo (MC) method are used to numerical approximations of the multidimensional integrals \cite{liu2008monte, robert2013monte}. In particular, the high fidelity model $f \left( \cdot \right)$ is replaced by the deep neural networks based surrogate model $\hat{f} \left( \cdot \right)$. And the integral for the statistical moments as well as the PDF can be consequently expressed in the form of the averaging operator using sampled values:

\begin{equation}
    \label{eq: 27}
    \begin{aligned}
    \boldsymbol{\mu}_{f} & = \lim_{N \rightarrow \infty} \frac{1}{N} \sum_{i=1}^{N} \hat{f} \left( \boldsymbol{x}_i \right) \\
    \boldsymbol{\sigma}_{f}^{2} & = \lim_{N \rightarrow \infty} \frac{1}{N-1} \sum_{i=1}^{N} \left(  \hat{f} \left( \boldsymbol{x}_i \right) - \boldsymbol{\mu}_{f} \right)^2 \\
    p_{f}(\boldsymbol{y}) & = \lim_{N \rightarrow \infty} \frac{1}{N} \sum_{i=1}^{N} \delta(\boldsymbol{y}-\hat{f}(\boldsymbol{x}))
    \end{aligned}
\end{equation}

The noteworthy fact is the combined use of the surrogate model and MC method in the context of the uncertainty quantification can usually reduce the computational burden by orders of magnitude. More importantly, the construction of a reliable and robust surrogate model is of critical significance throughout the uncertainty analysis. In the following section, we propose a deep neural networks based surrogate to approximate the mapping relationship between discretized high-dimensional random fields.

\section{Methodology}
\label{sec3}
The fundamental structure of the proposed surrogate model is based on a specific neural networks architecture, that is, the convolutional neural networks (CNNs) \cite{lecun1998gradient}. We build our surrogate in a hierarchical manner so that the model is more efficient in terms of training and deploying. Because we mainly focus on the regression problems of random fields, we name the model \textit{field regressor}. The rest of this section covers a guideline on the construction, training, and validation of a field regressor. 

\subsection{Model construction}
\label{sec31}
\begin{figure}[H]
\centering
\includegraphics[width=1.0\textwidth]{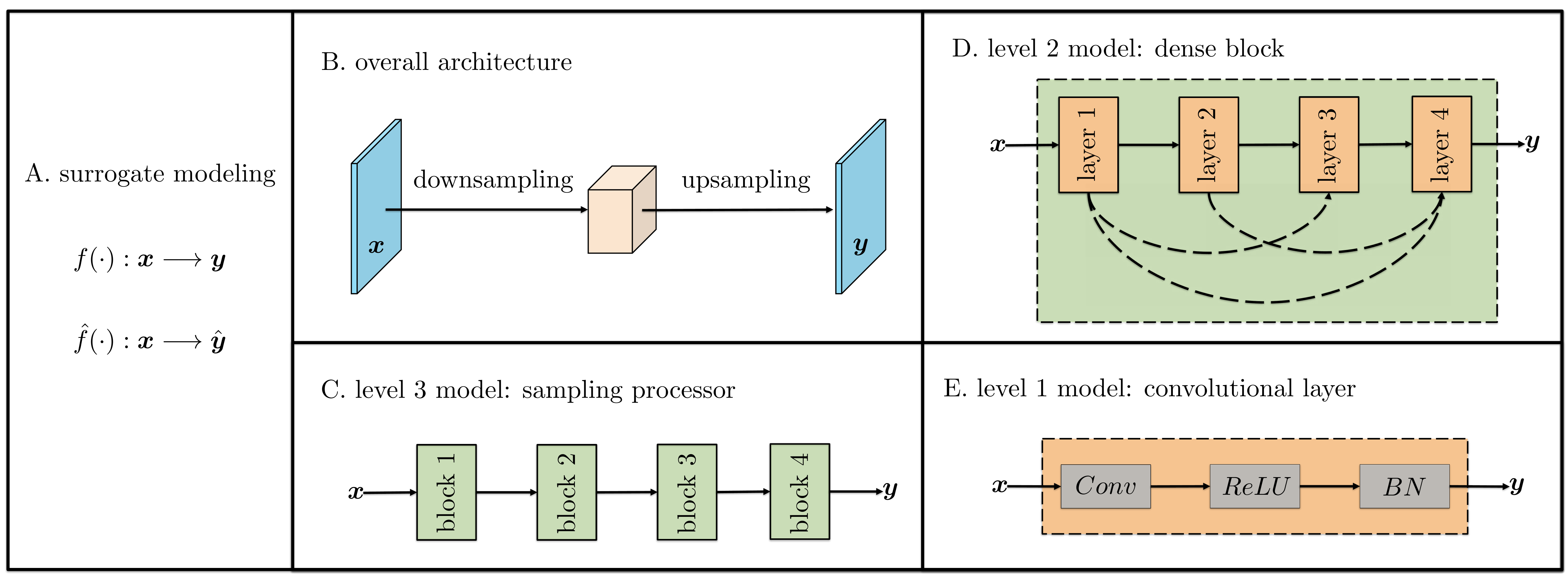}
\caption{Schematic diagram of field regressor} 
\label{fig: f2}
\end{figure}

Given the input data set $\boldsymbol{x}$ and the desired output data set $\boldsymbol{y}$, we aim to build a machine-learned model $\hat{f} \left( \cdot \right)$ such that $\hat{f} \left( \cdot \right)$ can approximate the input-output relationship $\boldsymbol{y} = f\left( \boldsymbol{x} \right)$. In the present case, $\boldsymbol{x}$ and $\boldsymbol{y}$ are discretized random fields. Let $\boldsymbol{x} \in \mathbb{R}^{H \times W \times C}$ and $\boldsymbol{y} \in \mathbb{R}^{H^{\prime} \times W^{\prime} \times C^{\prime}}$, where $H (H^{\prime})$ is the y-axis resolution, $W (W^{\prime})$ is the x-axis resolution, and $C (C^{\prime})$ denotes the number of random fields. The proposed surrogate $\hat{f} \left( \cdot \right)$ is essentially a hierarchical model that contains three different levels. It approximates the nonlinear regression function $f \left( \cdot \right)$ through a hierarchy of convolutional layers. \cref{fig: f2} summarizes the specifications for the architecture and assembling of the surrogate $\hat{f} \left( \cdot \right)$. We will investigate the detailed designs in a bottom-up manner, starting from level $1$ model.

\subsubsection{Level 1 model: convolutional layer}
\label{sec311}
A convolutional layer by its most prevailing definition is a composite function that has three mathematical operations \cite{dumoulin2016guide}. First, a convolutional layer performs element-wise multiplications between a subarray of the input field $\boldsymbol{x}$ and the kernel matrix $\mathbf{W} \in \mathbb{R}^{h \times w \times C}$ to obtain the so-called feature value at location $(m, n)$: 

\begin{equation}
    \label{eq: 31}
    \boldsymbol{\gamma}_{m, n} = (\boldsymbol{x} \star \mathbf{W})^{s}_{m, n} = \sum_{p=1}^{h} \sum_{q=1}^{w} \sum_{r=1}^{C} \mathbf{W}^{s}_{p, q, r} \cdot \boldsymbol{x}_{p+m,q+n,r}
\end{equation}

where $s$ denotes the kernel number, i.e. $s = 1, 2, \dots, C^{\prime}$. The computed output matrices are often called the feature maps, which contain the local correlation information about the input field. There are two other important parameters for the convolutional layer: the stride number $s$ and the padding number $p$. Specifically, $s$ represents the distance between two consecutive kernels and $p$ introduces zeros values to the borders of the input fields. The selection of $h$, $w$, $s$, and $p$ together determines the size of the output fields \cite{goodfellow2016deep}:

\begin{equation}
    \label{eq: 32}
    W^{\prime}=\frac{W-w+2 p}{s}+1 \quad \text{and} \quad H^{\prime}=\frac{H-h+2 p}{s}+1
\end{equation}

\begin{figure}[H]
\centering
\includegraphics[width=1.0\textwidth]{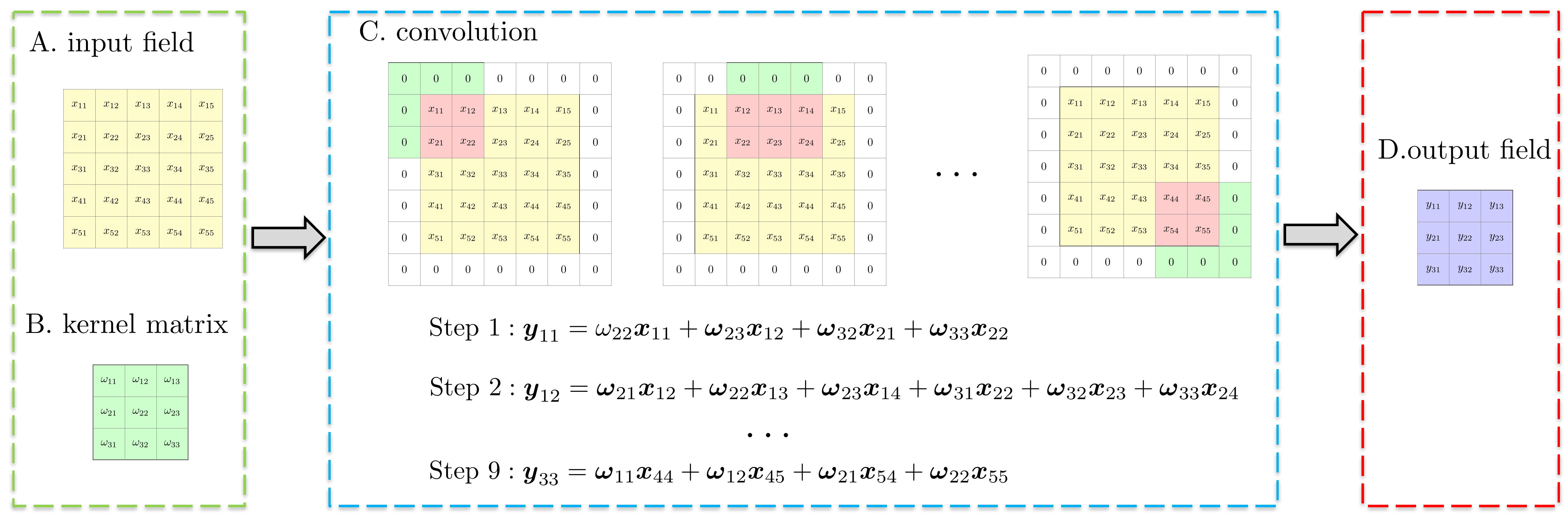}
\caption{Illustration example of the convolution arithmetic} 
\label{fig: f3}
\end{figure}

\cref{fig: f3} gives an example illustrating the convolution arithmetic. In this example, the input field $\boldsymbol{x} \in \mathbb{R}^{5 \times 5 \times 1}$, the kernel matrix $\mathbf{W} \in \mathbb{R}^{3 \times 3 \times 1}$, the stride number $s=2$ and the padding number $p=1$. Hence, the output field takes the form of $\boldsymbol{y} \in \mathbb{R}^{3 \times 3 \times 1}$.

Secondly, the output of the convolution flows over an element-wise nonlinear function that is often referred to as the activation function \cite{goodfellow2016deep}. Numerous types of activation functions exist in the literature and popular choice include the Sigmoid, the hyperbolic tangent or the rectified linear unit (ReLU). In conjunction with empirical knowledge, ReLU is implemented in the proposed scheme for two reasons. It has better gradient propagation property and analytical expressions of the ReLU function and its gradient are easy-to-implement as they only require basic addition and multiplication. An element-wise expression is given as:

\begin{equation}
    \label{eq: 33}
    h(\boldsymbol{\gamma}_{m, n}) = \max (\boldsymbol{\gamma}_{m, n}, \epsilon) \quad \text{and} \quad h^{\prime}(\boldsymbol{\gamma}_{m, n}) = \left\{\begin{array}{ll}{1} & \text{if} \quad {\boldsymbol{\gamma}_{m, n}>\epsilon} \\ {0} & \text{if} \quad {\boldsymbol{\gamma}_{m, n} \leq \epsilon}\end{array}\right.
\end{equation}

with $\epsilon$ denoting a small number, typically $0$. By manipulating the network architecture (e.g. number of layers, size of each layer, etc.), one can learn functions of arbitrary complexity.

Thirdly, we apply the batch normalization to the output of \cref{eq: 33}: 

\begin{equation}
    \label{eq: 34}
    \boldsymbol{\vartheta} = \frac{\alpha \left( h ( \boldsymbol{\gamma} ) - \mu_{\boldsymbol{x}} \right)}{\sqrt{\sigma_{\boldsymbol{x}}^{2}+\epsilon}} + \beta
\end{equation}

where $\alpha$ and $\beta$ are parameters to be learned, $\mu_{\boldsymbol{x}}$ and $\sigma_{\boldsymbol{x}}^{2}$ are sample mean and variance respectively. By means of \cref{eq: 34}, the internal covariate shift of the network architecture is minimized, allowing the data to propagate through a deeply structured neural network without distortion \cite{ioffe2015batch}.

\subsubsection{Level 2 model: dense block}
\label{sec312}
A quick summary, the level 1 model, i.e. the aforestated convolutional layer is structured as $\Upsilon(\cdot): BN \circ ReLU \circ Conv (\boldsymbol{x})$. Neural networks approximate the input-output relationship $f(\cdot)$ as:

\begin{equation}
    \label{eq: 35}
    \hat{f}(\boldsymbol{x}) = \Upsilon^{K} \circ \Upsilon^{K-1} \circ \ldots \Upsilon^{1}(\boldsymbol{x})
\end{equation}

where $K$ is the number of layers defined in the network architecture and $\circ$ is taken to be the composition operator with $\Upsilon^j, j \in\{1,2, \cdots, K\}$. Usually, CNNs take the output of the $(l-1)^{th}$ convolutional layer as the input to the $(l)^{th}$ layer:

\begin{equation}
    \label{eq: 36}
    \boldsymbol{x}_{l}=\boldsymbol{y}_{l-1}=\Upsilon_{l-1}\left(\boldsymbol{x}_{l-1}\right)
\end{equation}

Though simple and neat, recent research reveals such block design often encounters information degradation problems, which impede the training of CNNs \cite{glorot2010understanding}. To remedy this issue, distant connections among nonadjacent layers are established, where every single layer is fully connected with all the subsequent layers \cite{he2016deep, huang2017densely}. Consequently, the improved expression of function writes as:

\begin{equation}
    \label{eq: 37}
    \boldsymbol{x}_{l}=\boldsymbol{y}_{l-1}=\Upsilon_{l-1}\left(\left[\boldsymbol{x}_{0}, \boldsymbol{x}_{1}, \ldots \boldsymbol{x}_{l-1}\right]\right)
\end{equation}

In \cref{fig: f2}. (D), these augmented connections are depicted by the dashed lines. And a sequence of convolutional layers with nonadjacent layers is called dense block. It is assumed that each $\Upsilon_{i}(i=1,2, \ldots l-1)$ extracts $k$ feature maps. As a result, the input field for the $(l)^{th}$ layer has a size of $\mathbb{R}^{W_{l} \times H_{l} \times C_{l}}$, where $C_{l}=k_{0}+k(l-1)$. Noticeably, $k_{0}$ denotes the number of random fields of the first layer and $k$ is the growth rate.

\subsubsection{Level 3 model: sampling processor}
\label{sec313}
To improve the efficiency of $\hat{f} (\cdot)$ in terms of training, the high-dimensional input data $\boldsymbol{x}$ will first go through a downsampling processing, where feature maps produced by different convolutional layers are continually downsized to a coarser spatial resolution $H_{l}<H_{l-1} \,\, \& \,\, W_{l}<W_{l-1}$, resulting in a significant reduction of model parameters. This can be achieved either by controlling the stride number $s$ or using max pooling operation. In this paper, we choose to reduce the size of $\boldsymbol{x}$ by adopting non-unit stride so that sub-regional data relations can be better preserved \cite{dumoulin2016guide}. On the other hand, the dimension-reduced data will be projected back to the high dimension output space. Projection methods mainly fall into two categories, i.e. the deconvolution method and the interpolation method. We choose to resize the data by means of bicubic interpolation since the deconvolution method tends to produce checkerboard artifacts \cite{odena2016deconvolution}.  

\subsection{Model training}
\label{sec32}
\subsubsection{Regularized loss function}
\label{sec321}
Training a field regressor centers on optimizing the network parameters $\boldsymbol{\theta}$ in such a way that the discrepancy between model predictions $\hat{\boldsymbol{y}}_i$ and true observations $\boldsymbol{y}_i$ is minimized. For regression tasks, the discrepancy is typically measured by the mean squared error (MSE) loss. Given a training dataset $\mathcal{D} = \{ \boldsymbol{x}_i, \boldsymbol{y}_i \}_{i = 1}^{N}$ consisting of $n$ independent and identically distributed (i.i.d.) samples, the loss function is defined as:

\begin{equation}
    \label{eq: 38}
    \mathcal{L}\left(\boldsymbol{\theta}; \mathcal{D} \right)=\frac{1}{N} \sum_{i=1}^{N}\left\|\boldsymbol{y}_{i}-\hat{f}\left(\boldsymbol{x}_{i}, \boldsymbol{\theta}\right)\right\|^{2}
\end{equation}

In parctice, neural nets are notorious for overfitting. One usually resort to regularization techniques, penalizing $f^{'}(\cdot)$ by either promoting sparsity or driving weights to $0$. In this paper, $L_2$ norm is chosen to penalize the model complexity \cite{goodfellow2016deep}. The regularized MSE loss writes as:

\begin{equation}
    \label{eq: 39}
    \mathcal{L}\left(\boldsymbol{\theta}; \mathcal{D} \right)=\frac{1}{N} \sum_{i=1}^{N}\left\|\boldsymbol{y}_{i}-\hat{f}\left(\boldsymbol{x}_{i}, \boldsymbol{\theta}\right)\right\|^{2} + \lambda \sum_{j=1}^{K} \left\|\mathbf{W}^{(j)}\right\|_{2}^{2}
\end{equation}

where $K$ is the total number of layers, and $\lambda$ is a small constant that forces $f^{'}(\cdot)$ learning small weights (often termed as the weight decay), and $\mathbf{W}^{j}$ denotes all the parameters in the $j^{th}$ layer. For the field regressor, $\mathbf{W}^{j}$ includes the kernel weights utilized in the convolution operation, the scale and shift parameters in the batch normalization operation, and bicubic interpolation parameters used in the image resampling.

\subsubsection{Mini-batch stochastic gradient descent}
\label{sec322}
In order to compute the parameters $\boldsymbol{\theta}$, the gradient-based optimization is adopted as $f^{'}(\cdot)$ is a highly composite function involving a series of layers of nonlinear transformation \cite{rumelhart1988learning}. In particular, the Mini-batch stochastic gradient descent (mini-batch SGD) is implemented to solve this problem \cite{goodfellow2016deep}. The central idea is to approximate the negative gradient of the loss function by means of the expectation of a batch of samples:

\begin{equation}
    \label{eq: 310}
    \boldsymbol{\theta}_{k+1} \leftarrow \boldsymbol{\theta}_{k} - \eta_{k} \nabla_{\boldsymbol{\theta}} \mathcal{L}\left(\boldsymbol{\theta}; \mathcal{D}_{M} \right)
\end{equation}

where $\eta_{k}$ is the learning rate in the $k^{th}$ iteration and $\mathcal{L}\left(\boldsymbol{\theta}; \mathcal{D}_{M} \right)$ is the loss function evaluated using a mini-batch subset of samples $\mathcal{D}_{M} \subset \mathcal{D}$. Specifically, a mini-batch is initialized between $5$ and $500$ samples, chosen through a random manner and the Adaptive Moments (ADAM) optimization algorithm, a variant of the SGD methods, is selected to update the learnable parameters \cite{kingma2014adam}:

\begin{equation}
    \label{eq: 311}
    \boldsymbol{\theta}_{k+1} \leftarrow \boldsymbol{\theta}_{k}+\eta_{k} \frac{M_{k}}{1-\beta_{1}^{k}} /\left(\sqrt{\frac{V_{k}}{1-\beta_{2}^{k}}}+\epsilon\right)
\end{equation}

In \cref{eq: 311}, $\epsilon$ is a small number introduced to prevent $0$ denominator, and $M_{k}$ and $V_{k}$ are estimates of the mean and variance of the gradients, respectively. In ADAM, the update scheme of them are given as follows:

\begin{equation}
    \label{eq: 312}
    \begin{aligned}
    M_k &= \beta_1 M_{k-1} + (1 - \beta_1) \mathcal{L}\left(\boldsymbol{\theta}; \mathcal{D}_{M} \right) \\  
    V_k &= \beta_2 V_{k-1} + (1 - \beta_2) \mathcal{L}^2\left(\boldsymbol{\theta}; \mathcal{D}_{M} \right)  
    \end{aligned}
\end{equation}

Note that $M_{0}$ and $V_{0}$ are set to $0$ during the initialization. $\beta_1$ and $\beta_2$ are averaging parameters. In this work, values are configured as $\beta_1 = 0.9$ and $\beta_2 = 0.999$ following the suggestion in \cite{kingma2014adam}.

\subsection{Model validation}
\label{sec33}
To assess the performance of a fully trained model, $f^{'} \left( \cdot \right)$ is tested on a validation dataset $\mathcal{D} = \{ \boldsymbol{x_i}, \boldsymbol{y_i} \}_{i=1}^{M}$. More specifically, two performance indicators are selected to check whether $f^{'} \left( \cdot \right)$ provides a good approximation. First, root mean squared error (RMSE) is used to monitor the convergence history of model training:

\begin{equation}
\begin{aligned}
\label{eq: 313}
RMSE = \sqrt{\frac{1}{M} \sum_{i=1}^{M} || \hat{\boldsymbol{y}}_i - \boldsymbol{y}_i ||_2^2}
\end{aligned}
\end{equation}

Secondly, the coefficient of determination ($R^2$) is selected as the performance indicator. In statistics, $R^2$ estimates the proportion of variance in $\boldsymbol{y}$ that can be explained by $f^{'} \left( \cdot \right)$:

\begin{equation}
\begin{aligned}
\label{eq: 314}
R^2 = 1 - \frac{Var \left( \boldsymbol{y} | \boldsymbol{x} \right)}{Var \left( \boldsymbol{y} \right)} = 1 - \frac{\sum_{i=1}^{M} || \hat{\boldsymbol{y}}_i - \boldsymbol{y}_i ||_2^2}{\sum_{i=1}^{M} || \bar{\boldsymbol{y}}_i - \boldsymbol{y}_i ||_2^2}
\end{aligned}
\end{equation}

where $\bar{\boldsymbol{y}}_i$ is the mean value. From a mathematical perspective, $R^2$ takes on values between 0 and 1. When $R^2$ is close to one, it indicates $f^{'} \left( \cdot \right)$ fits the data well. On the other hand, $R^2 = 0$ means the model does not explain any variability in $\boldsymbol{y}$. Usually, $R^2 > 0.5$ reflects a significant relationship between $\boldsymbol{x}$ and $\boldsymbol{y}$.

\section{Case study}
\label{sec4}
\subsection{Problem setup}
\label{sec41}
\begin{figure}[H]
\centering
\includegraphics[width=1.0\textwidth]{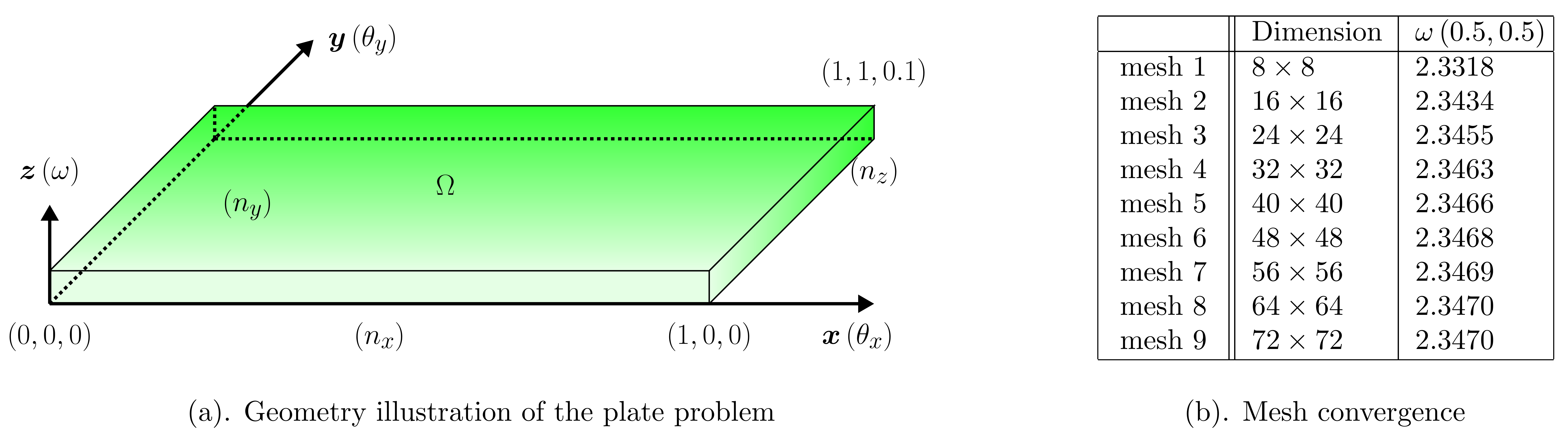}
\caption{Illustration of the case study}
\label{fig: f4}
\end{figure}

To demonstrate the effectiveness and efficiency of the proposed surrogate, we consider the following benchmark Mindlin–Reissner model on the unit square domain \cite{mindlin1951influence, reissner1945effect}:

\begin{equation}
\begin{aligned}
\label{eq: 41}
{-\operatorname{div} \mathbf{C} \varepsilon(\boldsymbol{\theta})-\gamma=0} & \quad  {\text { in } \quad  \Omega} \\ {-\operatorname{div} \gamma=f} & \quad  {\text { in } \quad \Omega} \\ {-\lambda t^{-2}(\boldsymbol{\nabla} w-\boldsymbol{\theta})} & \quad  {\text { in } \quad \Omega} \\ {\boldsymbol{\theta}=0, w=0} & \quad  {\text { on } \quad \partial \Omega}
\end{aligned}
\end{equation}

where $\Omega = [0, 1]^2 \subset \mathbb{R}^2$ is a smooth domain with a small thickness $t = 0.1$, $\mathbf{C}$ is the positive definite tensor denoting the bending moduli, $\varepsilon$ is the linear Green strain tensor, $\boldsymbol{\theta} = [\theta_x, \theta_y]$ represents the rotations of the surface, $w$ is the transverse displacement in z-direction, $\gamma$ denotes the scaled shear stresses, $f$ is the applied scaled transversal load, and $\lambda = E \kappa / 2(1+v)$ is the shear modulus with $E$ denoting the Young’s modulus, $v = 0.3$ denoting the Poisson ratio, and $\kappa = 5/6$ denoting the shear correction factor. Essentially, \cref{eq: 41} is a system of second order partial differential equations that can describe the bending behavior of a clamped plate loaded by a transverse force and \cref{fig: f4}. (a) shows the computational domain of our interest.

The objective of this benchmark problem is to build a surrogate model that can propagate and quantify the spatially-varying uncertainty associated with model inputs. The selected random input property is modeled by means of random field theory where its logarithm is a Gaussian random field:

\begin{equation}
\label{eq: 42}
\log \boldsymbol{x}(s) \sim \operatorname{GP}\left(m(s), k\left(s, s^{\prime}\right)\right)
\end{equation}

where $m(s)$ and $k\left(s, s^{\prime}\right)$ are the mean and covariance functions of the Gaussian process. In this case study, the mean function is defined to be zero and the exponentiated quadratic covariance function that is also known as the RBF kernel is adopted:

\begin{equation}
    \label{eq: 43}
    k\left(s, s^{\prime}\right)=\sigma^{2} \exp \left(-\frac{\left(s-s^{\prime}\right)^2}{2 \ell^2}\right)
\end{equation}

where the correlation length $l = 0.5$. We expand the scope of the case study by applying the proposed approach to the modeling of different mapping relationships, even though there is no obvious structure sharing between the input and output fields. Quantities of interest in different cases are summarized in \cref{table: t1}:

\begin{table}[h]
\centering
\begin{tabular}{l l l l}
\hline
\textbf{Case study} & \textbf{Inputs} & \textbf{Outputs} & \textbf{Surrogate modeling}\\
\hline
Case $1$ & $\boldsymbol{x} = E(s)$ & $\boldsymbol{y} = w(s)$ & $\hat{f}(\cdot): \mathbb{R}^{4096} \rightarrow \mathbb{R}^{4096}$\\
Case $2$ & $\boldsymbol{x} = E(s)$ & $\boldsymbol{y} = [\sigma_{v}(s), \tau_{\max }(s), \tau_{x y}(s)]^{\dagger}$ & $\hat{f}(\cdot): \mathbb{R}^{4096} \rightarrow \mathbb{R}^{12288}$ \\
Case $3$ & $\boldsymbol{x} = [ E(s), f(s) ]$ & $\boldsymbol{y} = [w(s), \sigma_{v}(s)]$ & $\hat{f}(\cdot): \mathbb{R}^{8192} \rightarrow \mathbb{R}^{8192}$ \\
\hline
\end{tabular}
\begin{tablenotes}
\item $\dagger$. $\sigma_{v}$ is the Von Mises stress, $\tau_{max}$ denotes the maximum shear stress, and $\tau_{xy}$ is the shear stress. They can be obtained by evaluating the quadrature points for the stress-strain relationship.
\end{tablenotes}
\caption{Summary of the case study analysis}
\label{table: t1}
\end{table}

\subsection{Training, testing, and validation datasets}
\label{sec42}
To obtain input-output data to construct the deep convolutional neural networks, we solve the forward PDEs using the finite element method (FEM) \cite{hughes2012finite}. After the convergence study (See \cref{fig: f4}. (b)), the unit square domain is discretized by $64 \times 64$ quadrilateral isoparametric element (Q4) elements, ensuring the accuracy of the FE solution. Hence, the number of degrees of freedom of the high-fidelity FE model is $12675$. The FEM writes the weak form of equilibrium equations (\cref{eq: 41}) of the system into a set of algebraic equations, and solves the whole set by the Newton-Raphson method. The input to the FEM solver is the discretized representation of the random field(fields). For instance, each sample $\boldsymbol{x}_i \in \mathbb{R}^{64 \times 64 \times 2}$ in the case $3$ as the input covers the material and the load. The last column of \cref{table: t1} shows the dimensionality of the input-output relationship of each case.

There are several sampling schemes available to generate data that includes three sets: the training set, testing set, and validation set. We use the Latin hypercube sampling method to generate samples from the prescribed probability distributions. Specifically, we generate $1024$ input samples at which the high-fidelity model is evaluated, yielding the training dataset $\mathcal{D}_{train} = \{ \boldsymbol{x}_i, \boldsymbol{y}_i \}_{i=1}^{1024}$. And we randomly sample another $200$ inputs for the testing $\mathcal{D}_{test} = \{ \boldsymbol{x}_i, \boldsymbol{y}_i \}_{i=1}^{200}$. The reference statistics of the uncertainty analysis is computed by the Monte Carlo method using $10^5$ samples. Once the constructed network is optimally trained, we carry out the uncertainty quantification on the easy-to-evaluate surrogate instead of the computationally intensive finite element model with the same input snapshots.
 
The surrogate modeling algorithm is implemented in the open source machine learning library TensorFlow and the scripts are tested on a single NVIDIA GeForce GTX 1080 Ti X GPU. For the purpose of reproducibility, the code to replicate the optimization results in this work will be made available at \url{https://xihaier.github.io} upon publication of this manuscript.

\subsection{Results}
\label{sec43}
\subsubsection{Case 1: one-to-one mapping}
\label{sec431}
The first case concerns the use of field regressor to learn one-to-one mapping. Let the input be the material field, i.e. the young modulus and the output be the displacement field. The model inputs $\boldsymbol{x} \in \mathbb{R}^{4096}$ are discretized random field realizations and the model outputs $\boldsymbol{y} \in \mathbb{R}^{4096}$ are the transverse deformation at the finite element centers. The uncertainty analysis of the target system with high-dimensional stochastic inputs covers two parts. First, we have to train a surrogate model $\hat{f} : \mathbb{R}^{4096} \times[0,1]^{2} \rightarrow \mathbb{R}^{4096}$, which can accurately map the snapshot of the material field to the displacement field governed by the PDEs. Next, the trained surrogate is applied to the uncertainty quantification tasks, computing statistical moments and PDFs.

\begin{figure}[H]
\centering
\includegraphics[width=0.97\textwidth]{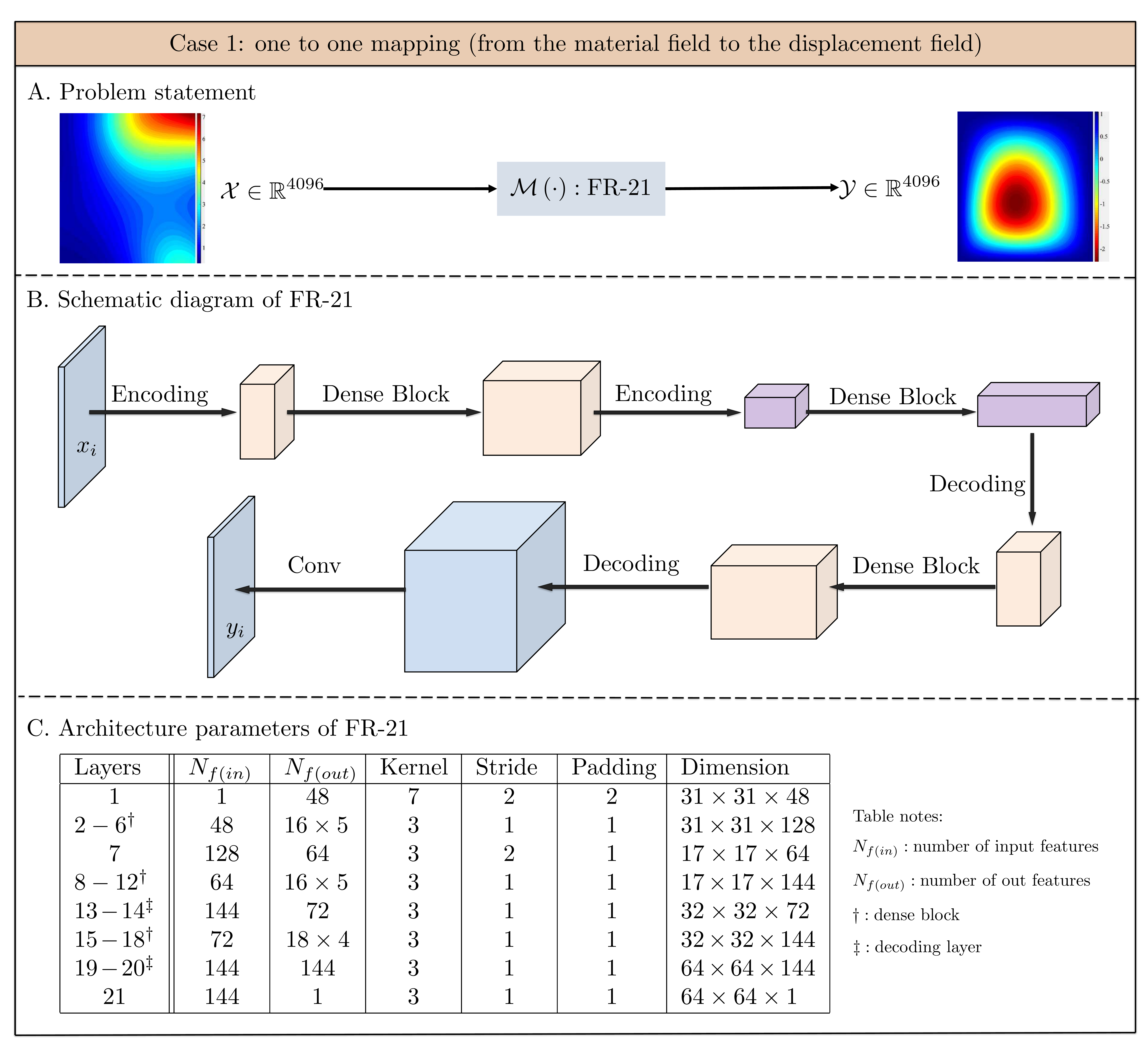}
\caption{Neural networks architecture design and parameterization of FR-21} 
\label{fig: f5}
\end{figure}

\textbf{Network architecture.} After extensive hyperparameter and network architecture search, the field regressor is built using 21 convolutional layers and this configuration is hence referred to as FR-21. Specifically, the dimension of each random field sample has been reduced twice by the downsampling processor $64 \times 64 \rightarrow 31 \times 31 \rightarrow 17 \times 17$. There are three dense blocks within FR-21. They share the same kernel size, stride number, and padding number while have different growth number and depth. \cref{fig: f5} gives a general summary of the one-to-one mapping problem, graphic illustration of the data flow of FR-21, and the details of these network configuration parameters.

To learn the high-dimensional input-output relationship, the gradient-based optimizer Adam is selected. We train the surrogate model (FR-21) with training data size varies from $64$ to $1024$. The initial learning rate is set to $\eta_0 = 0.005$ and the step decay learning strategy (often known as annealing) is adopted to gradually lower the learning rate during training. The default value $\eta_0$ is reduced by a annealing rate $\zeta = 0.75$ every $20$ epochs and the model trained with $500$ epochs in total. For regularization, the weight decay is initialized to $\lambda = 7 \times 10^{-6}$ and the batch size is $8$.

\begin{figure}[H]
\centering
\includegraphics[width=1.0\textwidth]{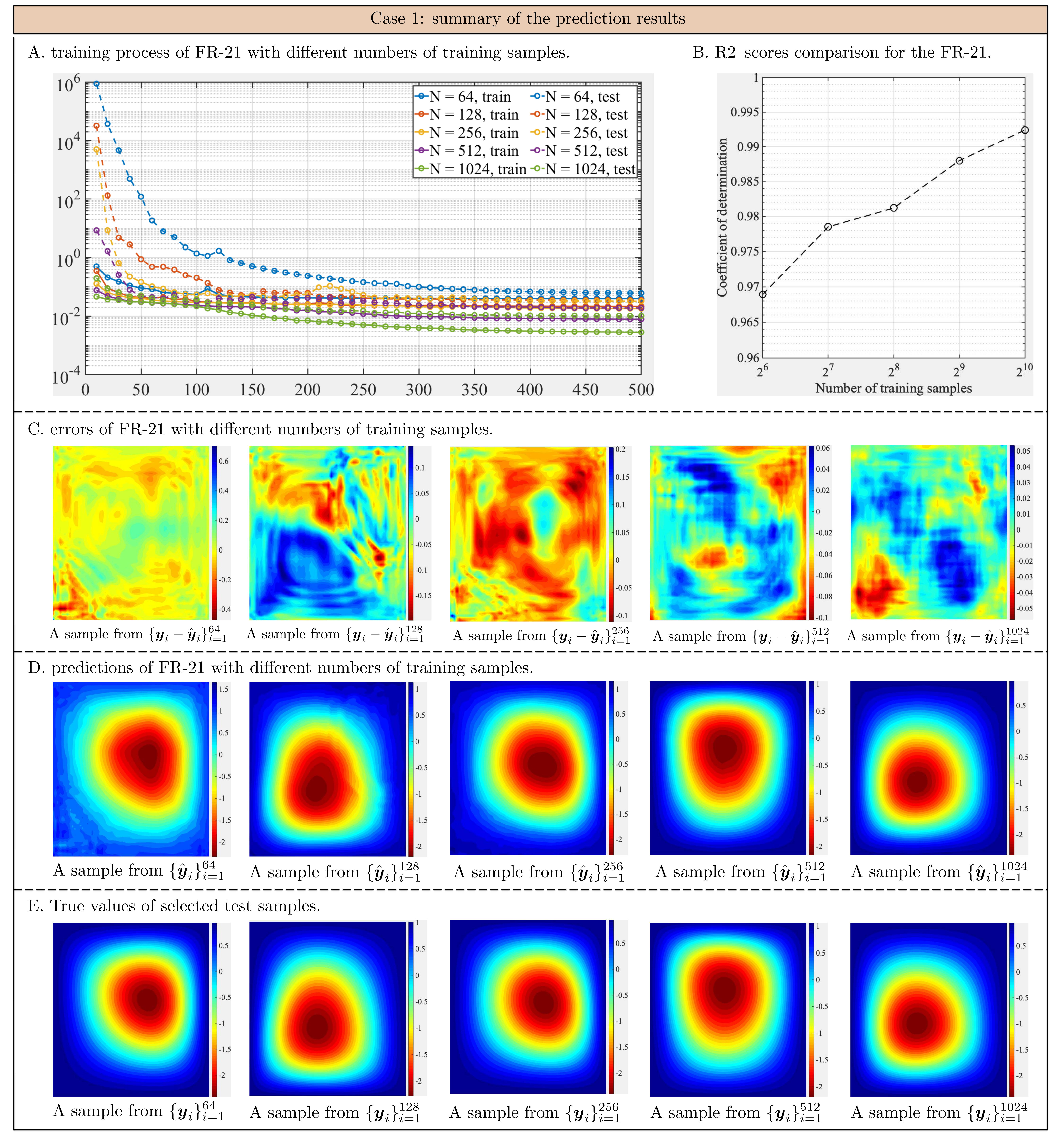}
\caption{Training and prediction performance of FR-21} 
\label{fig: f6}
\end{figure}

\textbf{Optimization results.} \cref{fig: f6} (A) plots the training process in terms of the RMSE value, which is recorded every $20$ epochs. The rapidness and steadiness of the convergence of the FR-21 can be seen from the results. In particular, after around $300$ epochs of the Adam optimizer, an approximately stabilized solution is achieved. The optimization algorithm converges faster with more training samples and the RMSE value of the testing dataset reduces as the training size increases: $64 \rightarrow \dots \rightarrow 1024$. Meanwhile, the coefficient of determination (also known as the $R^2$-score) is evaluated using 200 test samples with training dataset sizes. Note that the trained FR-21 shows a remarkable generalization ability where the $R^2$-score is more than $0.9$ in all the trials. In \cref{fig: f6} (B), the model achieves a relatively high $R^2$-score of $0.967$ with only $64$ training samples and the $R^2$-score gets close to $0.99$ when increasing the training dataset to $512$ samples, which indicates that the predicted output fields by FR-21 match the ground truth computed by the FEM very well. \cref{fig: f6} (C) show some of the predicted results. The errors of the predicted solutions of FR-21 are compared with respect to the FE solutions. Note that the results representing the output fields have been standardized so that a more clear comparison can be carried out between FR-21 with different training data sizes. For visualization purpose, we randomly selected a sample from the testing dataset $\mathcal{D}_{test} = \{ \boldsymbol{x}_i, \boldsymbol{y}_i \}_{i=1}^{200}$. Remarkably, the propsoed FR-21 is capable of providing good predictions. Even when using only $64$ training samples, the model is able to quickly capture the landscape of the target output field. The prediction quality can be significantly improved as more samples are put into the training process.

\textbf{Uncertainty analysis results.} The uncertainty analysis results predicted by the FR-21 is verified against the solution computed by the FEM solver. The number of training samples is set to $N = 1024$, and the number of testing samples is set to $M = 1 \times 10^5$. \cref{fig: f6} (A) and (B) show the the mean and variance of the displacement field, respectively. It can be seen that FR-21 achieves smooth contour estimation that is close to the Monte Carlo result. The maximum error of the estimate measured in a relative error form is less than $10 \%$. Meanwhile, we randomly select two points out of the domain and show the PDFs, respectively. The unbounded kernel smoothing function is used to estimate the PDFs, where the type of kernel smoother is set to normal. \cref{fig: f6} (C) compares the PDFs at two locations from FR-21 and FEM solver. As desired, the PDFs obtained by the surrogate model at two randomly selected locations are almost identical to the reference solution.

\begin{figure}[H]
\centering
\includegraphics[width=1.0\textwidth]{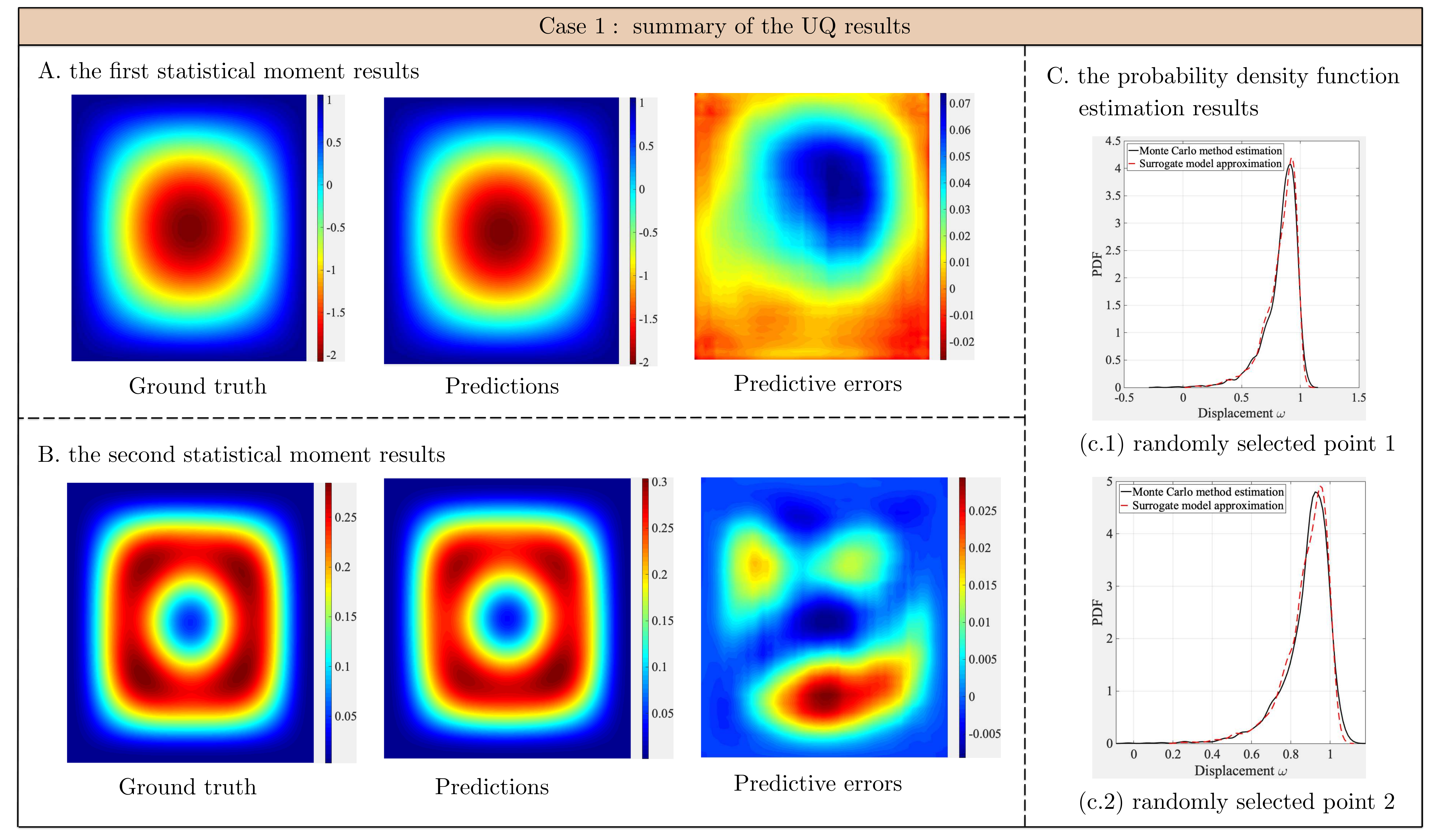}
\caption{Uncertainty quantification results of FR-21} 
\label{fig: f7}
\end{figure}

\subsubsection{Case 2: one-to-many mapping}
\label{sec432}
In this case, the field regressor is generalized to the surrogate learning of one-to-many mapping. The input field is the young modulus $\boldsymbol{x}=E(s)$, and the output fields are stress fields that essentially are functions of displacement fields $\boldsymbol{y}=\left[\sigma_{v}(s), \tau_{\max }(s), \tau_{x y}(s)\right]$ (See \cref{table: t1} for details). The objective is to build a surrogate model that can predict three discretized random fields simultaneously from one input random field $\hat{f} : \mathbb{R}^{64 \times 64} \times[0,1]^{2} \rightarrow \mathbb{R}^{64 \times 64 \times 3}$.

\begin{figure}[H]
\centering
\includegraphics[width=1.0\textwidth]{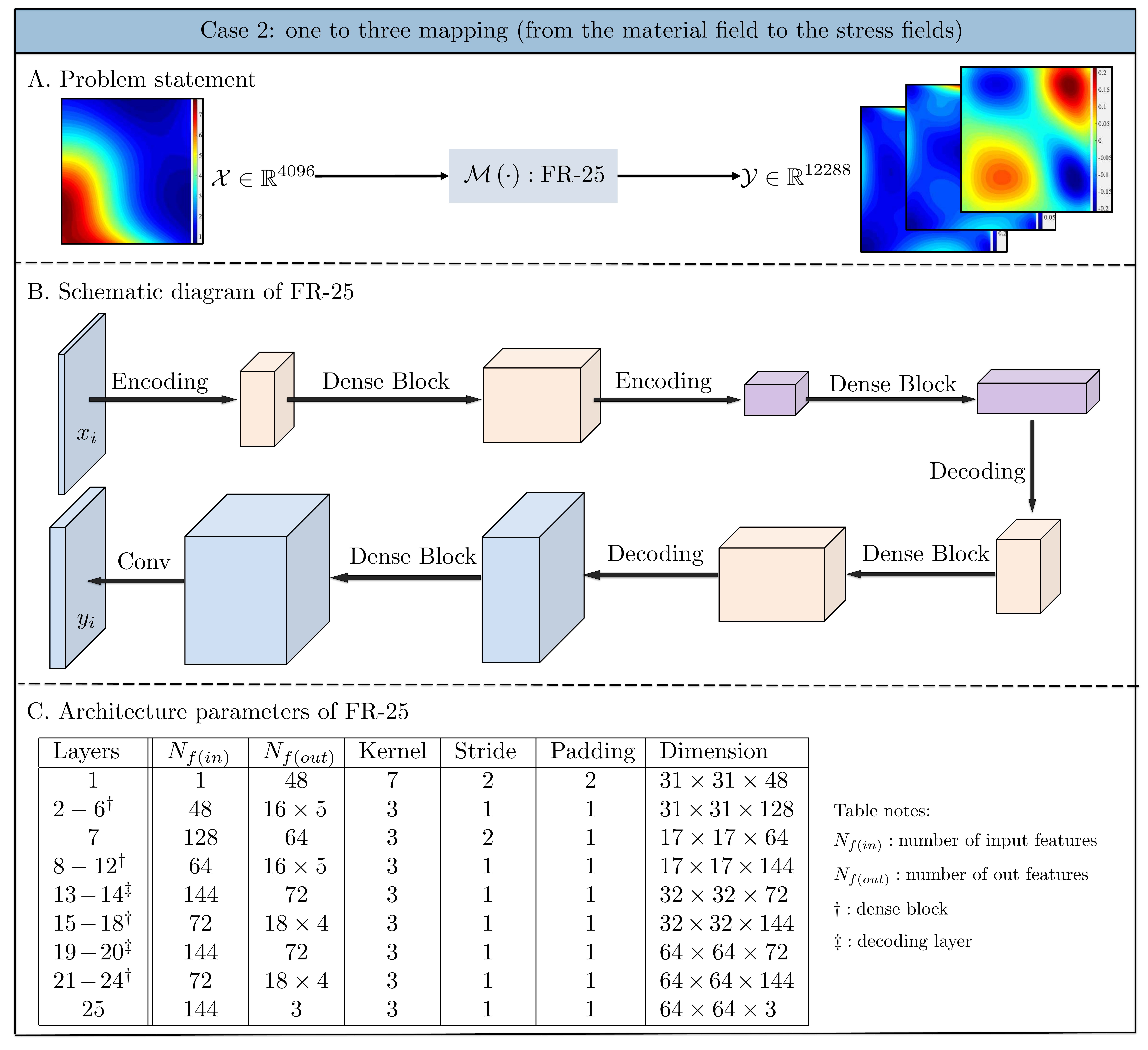}
\caption{Neural networks architecture design and parameterization of FR-25} 
\label{fig: f8}
\end{figure}

\textbf{Network architecture.} The surrogate model is derived using 25 convolutional layers in total, and hence is referred to as FR-25. There are four dense blocks in the FR-25. In particular, we build a dense block before the last convolutional layer, ensuring that sufficient information has been extracted to construct three independent output fields. The number of feature maps increases from $72$ before the block to $144$ after a set of block layers $21-24$. \cref{fig: f8} gives a graphic representation of the one-to-many mapping problem $\mathbb{R}^{4096} \rightarrow \mathbb{R}^{12288}$ and summarizes the detailed designs of FR-25.

\begin{figure}[H]
\centering
\includegraphics[width=1.0\textwidth]{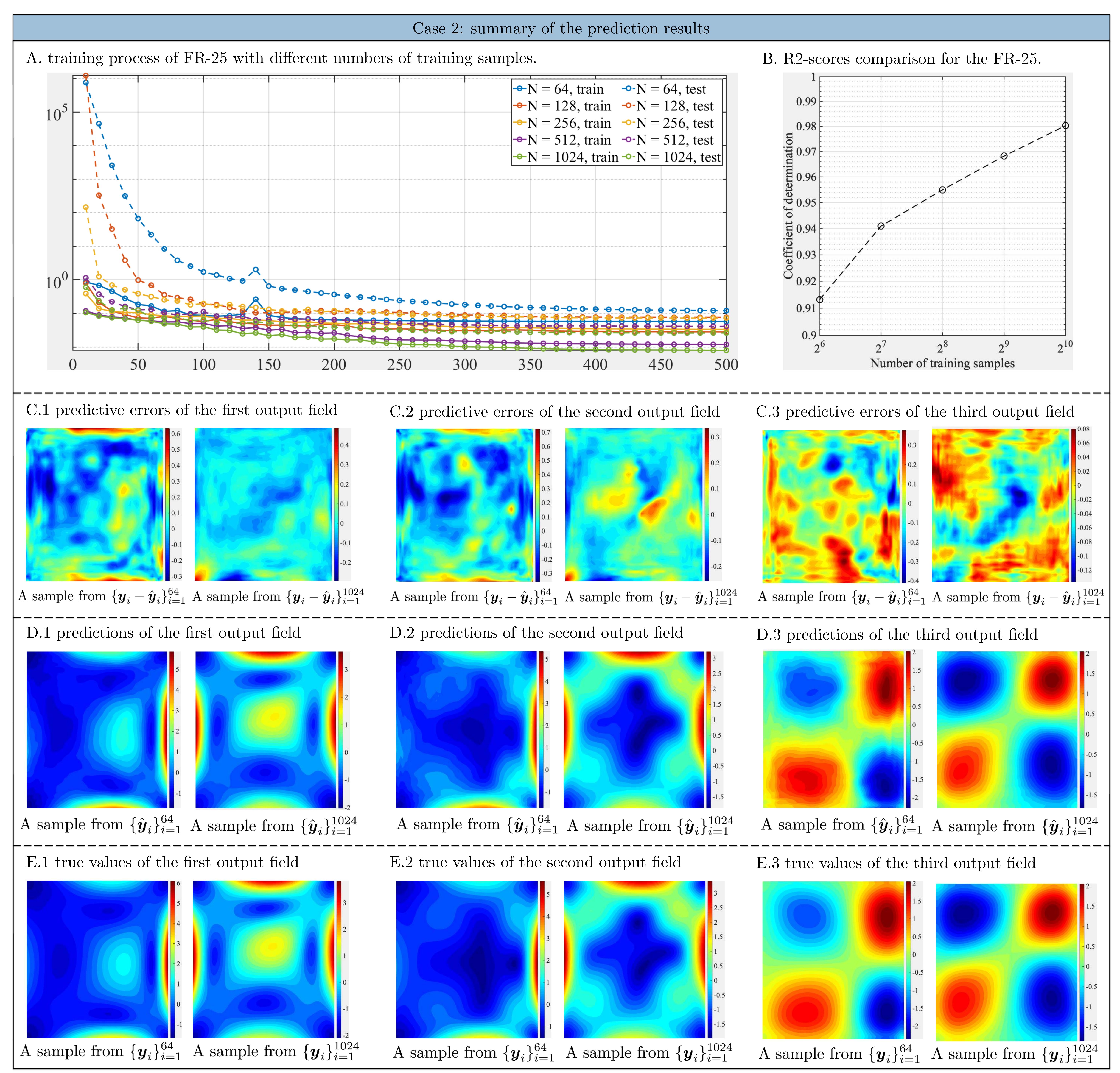}
\caption{Training and prediction performance of FR-25} 
\label{fig: f9}
\end{figure}

\textbf{Optimization results.} The surrogate model is trained with the same optimization configuration stated in \cref{sec431}. \cref{fig: f9} (A) confirms the learning effectiveness of our proposed surrogate model. The Adam optimizer reaches stabilized state after $350$ epochs. The objective function minimizes faster with more training data, and the RMSE value at epoch $500$ increases when the training size decreases. \cref{fig: f9} (B) shows the $R^2$-score for each training trial. Note that the $R^2$-score is evaluated using all the predicted fields, implying each sample $\hat{\boldsymbol{y}}_{i}$ in \cref{eq: 314} belongs to the real-valued vector space $\mathbb{R}^{64 \times 64 \times 3}$. The results show that the surrogate, FR-25, provides an accurate approximation of the high-dimensional input-output relationship even with a few samples (e.g. 64 samples). Specifically, these $R^2$-score values indicate the majority of the observed variation (approximately 90 percents) can be explained by the model. We present the predicted results using $64$ and $1024$ training samples in \cref{fig: f9} (C), respectively. The estimated stress fields via the FR-25 closely resemble the landscape of the ground truth data when only $64$ samples are available for the model training. And the prediction accuracy improves substantially when the training sample size increases to $1024$. Noticebly, the preiction resutls of the shear stress field $\tau_{xy}(s)$ is relatively less smoother, especially when the training sample size is small. This is due to the weights assigned to each output field. Without any prior knowledge, the intuitive assignment is the use of uniform weights. The different performances regarding different output fields are largely alleviated as we increase the training samples, and the predictions become very close to the reference solutions, where the relative maximum error of $\sigma_{v}(s)$, $\tau_{max}(s)$, and $\tau_{xy}(s)$ are in the same range. 

\begin{figure}[H]
\centering
\includegraphics[width=1.0\textwidth]{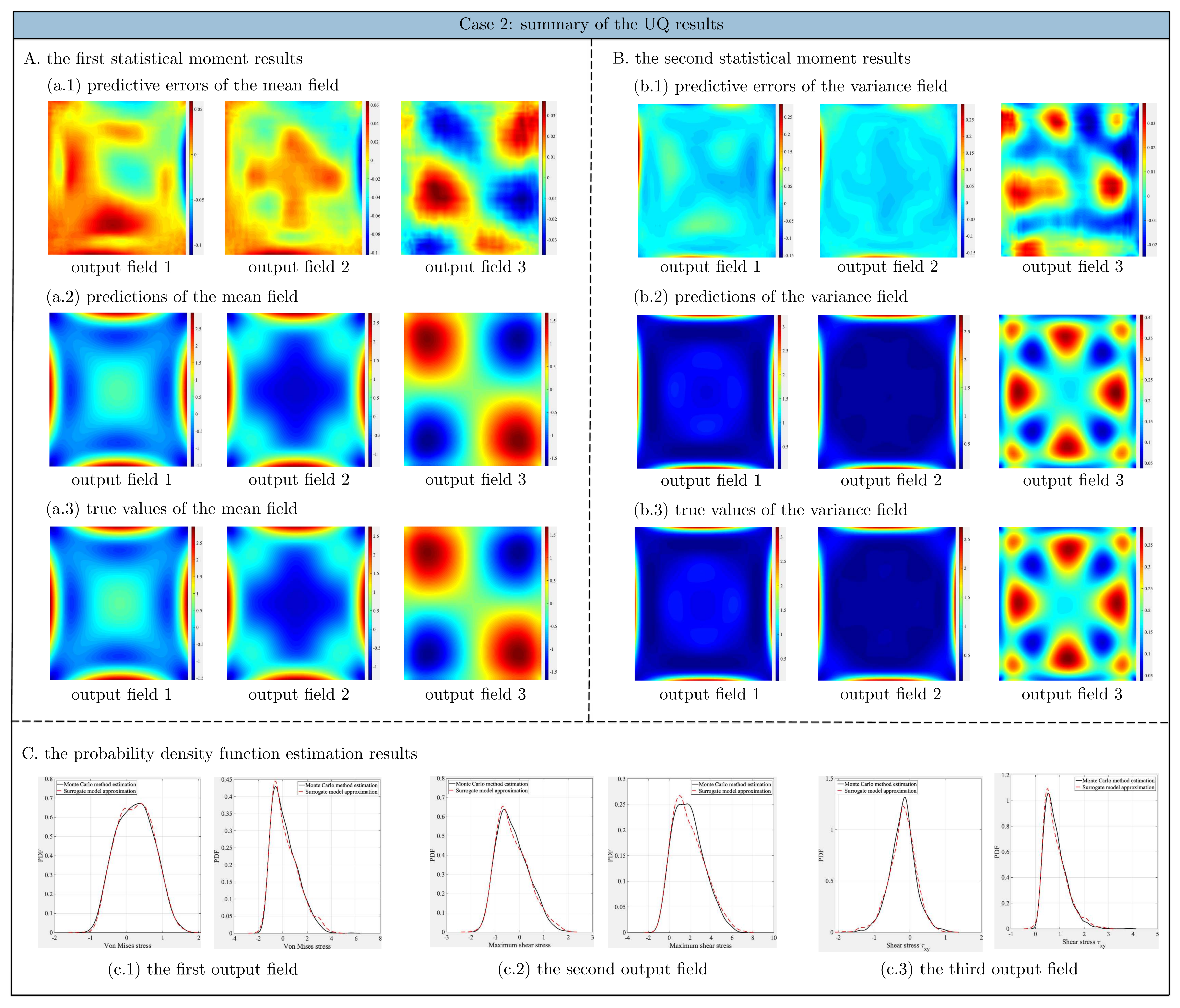}
\caption{Uncertainty quantification results of FR-25} 
\label{fig: f10}
\end{figure}

\textbf{Uncertainty analysis results.} The predictive performance of the FR-25 to the machine learning with a small dataset seen in this case is a strong indication of the effectiveness and robustness of using the proposed surrogate modeling approach for uncertainty analysis. A number of $1 \times 10^5$ independent and identically distributed samples have been generated for the computation of the reference statistics using the Monte Carlo method. With respect to the PDFs, we once again randomly select $2$ points from each output field and use kernel methods to estimate the corresponding PDF. For the mean field prediction, the maximum error is less than $5 \%$ measured in a relative error sense and the error centers around the extreme value region, for instance, the central cross region of the $\tau_{max}(s)$ field (See \cref{fig: f10} (A)). For the mean variance prediction, the maximum error is distributed on the boundary lines in terms of the prediction of $\sigma_{v}(s)$ and $\tau_{max}(s)$ and is overlapped with the extreme value region in the case of $\tau_{xy}(s)$ (See \cref{fig: f10} (B)). The PDFs obtained by the surrogate model at two randomly selected locations are almost identical to the reference solution. \cref{fig: f10} (C) shows the randomly selected points have non-gaussian distributions of the $\sigma_{v}(s)$, $\tau_{max}(s)$, and $\tau_{xy}(s)$ value. And the FR-25 can capture different PDFs with remarkably high precision.

\subsubsection{Case 3: many-to-many mapping}
\label{sec433}
The third and final case study investigates the efficacy of the proposed surrogate modeling approach for many-to-many mapping problems. We consider the boundary value problem defined by the elliptic PDEs (\cref{eq: 41}) with multiple random input fields. In particular, the material $E(s)$ and loading $f(s)$ properties are considered to vary spatially where the correlated variation has the same analytical form (\cref{eq: 42}). Subsequently, we are interested in characterizing the spatial variability induced effects on the structural performances that are denoted by the displacement $w(s)$ and stress $\sigma_{v}(s)$. The input-output relationship satisfies the high-dimensional mapping $\hat{f} : \mathbb{R}^{64 \times 64 \times 2} \times[0,1]^{2} \rightarrow \mathbb{R}^{64 \times 64 \times 2}$ on the unit square domain.

\begin{figure}[H]
\centering
\includegraphics[width=1.0\textwidth]{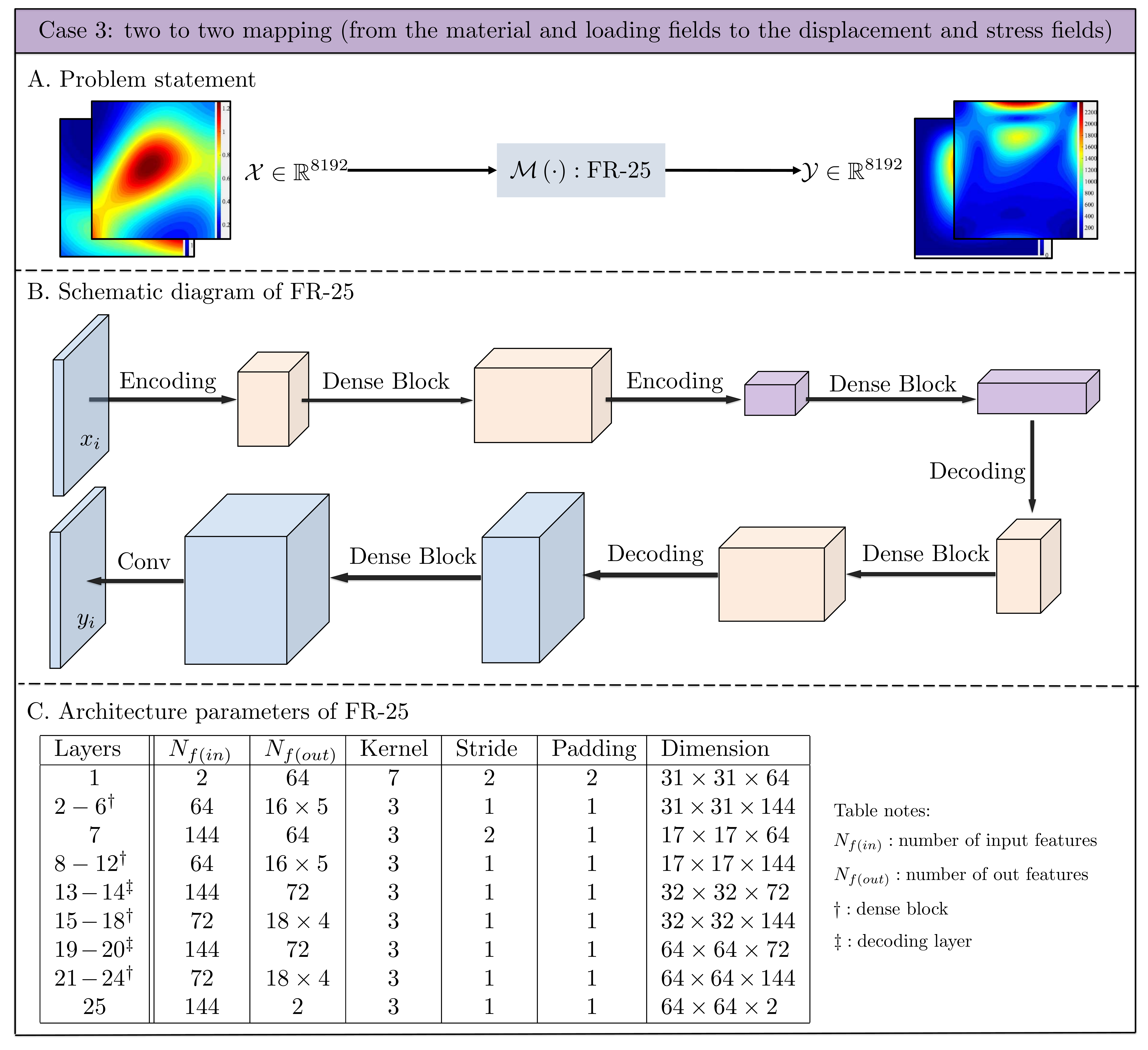}
\caption{Neural networks architecture design and parameterization of FR-25} 
\label{fig: f11}
\end{figure}

\textbf{Network architecture.} Similar to the previous case, the deep neural networks based surrogate model is configured with 25 convolutional layers. Since the model input $\boldsymbol{x}$ covers two independent random fields that are physically irrelevant, more features have been extracted and stacked by the end of the first layer, where the number of input and output features is $2$ and $64$, respectively. As displayed in the table of \cref{fig: f11}, a total of $4$ dense blocks have been constructed on account for the sufficiency of the reconstruction of the output fields. The concept of the many-to-many mapping problem is illustrated in \cref{fig: f11} (A) and a graphic diagram of the FR-25 in the present case is given in \cref{fig: f11} (B).

\begin{figure}[H]
\centering
\includegraphics[width=1.0\textwidth]{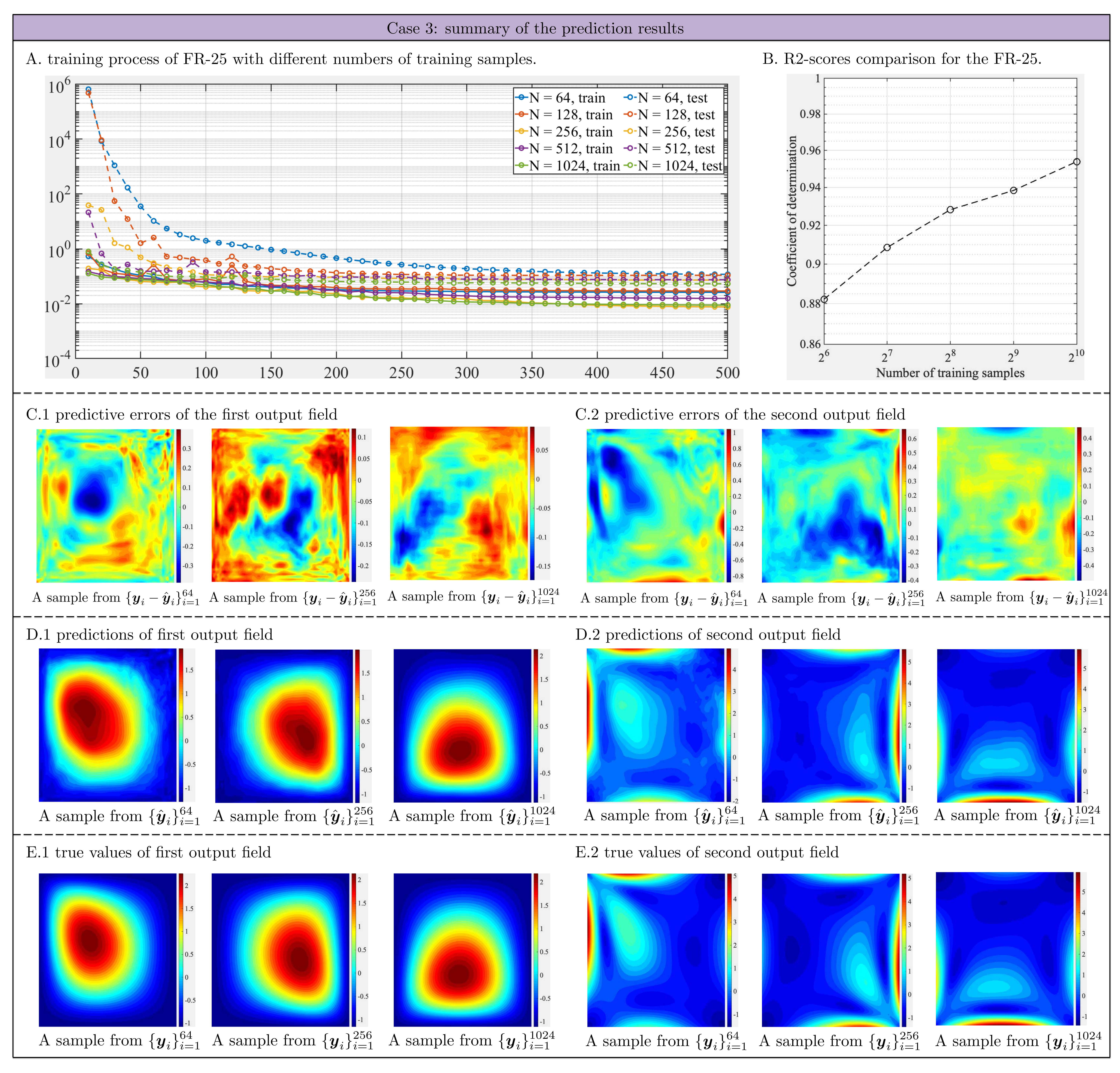}
\caption{Training and prediction performance of FR-25} 
\label{fig: f12}
\end{figure}

\textbf{Optimization results.} \cref{fig: f12} (A) shows the optimization history of the model learning with different size of training samples. The strength of the proposed surrogate in terms of training is verified as the objective function decreases dramatically within the first $200$ epochs. In comparing the different trials of this case study, again, it can be seen that an increase in the training sample size leads to a better-optimized result. Indeed, the $R^2$-score illustrates the importance of providing sufficient training samples, where the coefficient of determination value is relatively lower compared to the second case with the same amount of training samples, with the $R^2$-score of the trial using $64$ samples slightly smaller than $0.9$. This is due to the fact that the complexity of this case is higher than the second one. Potential improvement strategy includes altering the architecture design of the field regressor, for instant, extracting initial features in a separate manner rather than the joined style, that is,

\begin{equation}
    \label{eq: 44}
    \begin{aligned}
    & \text{Joint extraction: } && 64 \times 64 \times 2 &&& \rightarrow &&&& 31 \times 31 \times 64 \\ 
    & \text{Separate extraction: } && 64 \times 64 \times 2 = \begin{array}{ll} 64 \times 64 \times 1 \\ 64 \times 64 \times 1 \end{array} &&& \rightarrow &&&& \begin{array}{ll} 31 \times 31 \times 32 \\ 31 \times 31 \times 32 \end{array} = 31 \times 31 \times 64
    \end{aligned}
\end{equation}

For the purpose of easy comparison, we attempted to use a united design in all cases. The accuracy and consistency of the predicted results of the proposed approach can once again be seen from \cref{fig: f12} (C), (D), and (E). In particular, the precision observed in the selected trials of this case study approximately equivalent to those seen in the previous cases, and thus indicates the robustness and strong potential of the proposed surrogate modeling approach for solving a wide variety of problems of interest.

\begin{figure}[H]
\centering
\includegraphics[width=1.0\textwidth]{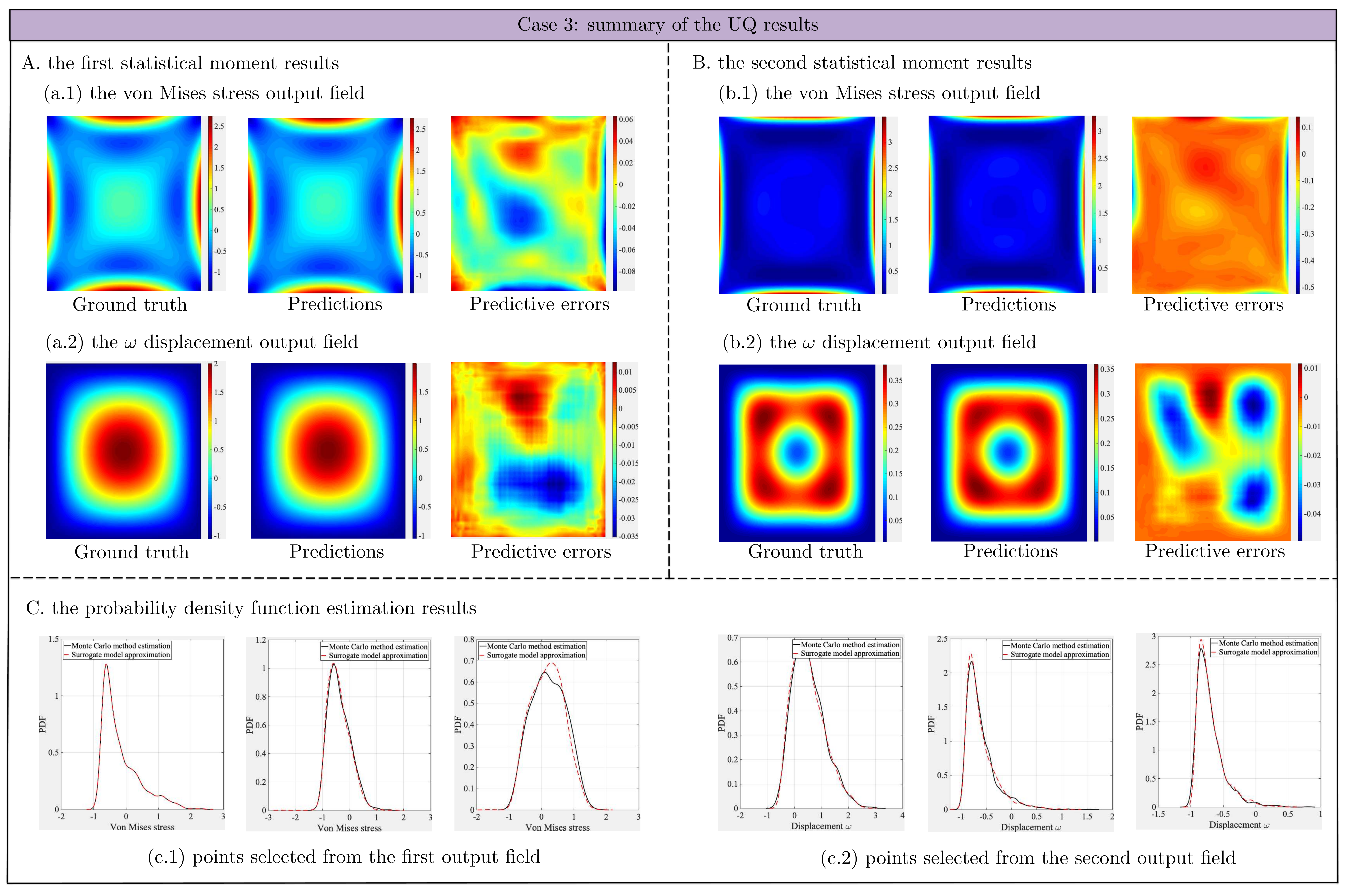}
\caption{Uncertainty quantification results of FR-25} 
\label{fig: f13}
\end{figure}

\textbf{Uncertainty analysis results.} \cref{fig: f13} first investigates the prediction performance of the mean and variance fields by means of FR-25. The reference solution is obtained by numerically solving the governing SPDEs with the finite element scheme $10^{5}$ times. \cref{fig: f13} (A) and (B) present the good agreement between the surrogate modeling and high-fidelity FE results for the statistics of the target output fields. \cref{fig: f13} (C) also depicts the PDFs of two output fields at three specific locations. It can be seen that our field regressor is able to accurately approximate the probabilistic response. This once again illustrates the robustness and efficiency of the proposed field regressor, i.e., using $1024$ samples is sufficient to capture most of the stochasticity in this $\mathbb{R}^{8192} \rightarrow \mathbb{R}^{8192}$ SPDEs.

\section{Conclusion}
\label{sec5}
This paper presents a deep neural network approach for uncertainty analysis of elliptic systems with spatially varying properties. Note that conventional random variable representation with uniform distribution of variance cannot consider the effect of spatial randomness notwithstanding the fact that spatial variability is naturally associated with a wide range of engineering systems. In the present study, we describe the correlation that exists in the uncertain spatial distribution by means of random field theory. After spatial discretization, the stochastic parameter space in terms of the modeling of the input-output relationship is high-dimensional. To overcome this high-dimensional stochasticity issue, we introduce a hierarchical surrogate model named the field regressor utilizing convolutional neural networks as the model basis. Compared to conventional approaches, our field regressor provides a direct approximation of the intrinsic input/output relationships without mapping the high-dimensional data to a lower dimension. 

Specifically, novelties of the proposed field regressor can be outlined in three perspectives. First, information flowing over the model is gradually downsized to a stack of feature maps. Model-estimated output fields are then reconstructed from extracted features. Such a multi-scaling data mechanism hereby advances the training process as fewer model parameters are required. Secondly, short connections between nonadjacent convolutional layers are introduced to the basic network architecture, allowing previously extracted features to be reused in the subsequent layers. This property favors the machine learning process as a deeply structured architecture becomes feasibly trainable and numerically stable. Thirdly, we use bicubic resizing and convolution operation in the upsampling process. Consequently, detailed variation patterns in the high-dimensional data can be better preserved.

The effectiveness and efficiency of the proposed field regressor is demonstrated on a benchmark UQ problem. It is observed that the field regressor is capable of directly inferring a variety of high-dimensional mapping relationships including one-to-one ($\mathbb{R}^{4096} \rightarrow \mathbb{R}^{4096}$), one-to-many ($\mathbb{R}^{4096} \rightarrow \mathbb{R}^{12288}$), and many-to-many ($\mathbb{R}^{8192} \rightarrow \mathbb{R}^{8192}$) mappings. Even when there is no obvious functional relationship between the data structures of the input and output fields, the proposed field regressor can accurately capture the intrinsic mapping relationships. The use of the field regressor greatly accelerates the uncertainty analysis, i.e. propagating the random field represented uncertainty to the output quantities of interest.

It should be noted that the current surrogate is merely built through data without integrating any physics into the neural network architecture. Possible improvement strategies of the model performance may include utilizing well-established physical interpretations during the architecture design process of the surrogate. For instance, building a new neural network as a representation of governing equations or engineering constraints in addition to the current field regressor will be helpful. In the future, besides applying the field regressor to other potential engineering applications with high-dimensional data, the current surrogate modeling technique will be extended to spatiotemporal problems.

\acknowledgments
X.L. acknowledges the NatHaz Modeling Laboratory at the University of Notre Dame where this work was performed. This research effort was supported in part by the National Science Foundation through Grant  No. 1520817 and No. 1612843. The codes and data used in this work will be made available at \url{https://xihaier.github.io/} upon publication of this manuscript.

\bibliography{manuscript.bib}


%
%




\end{document}